\documentclass[]{spie}  %>>> use for US letter paper
\newcommand{\K}{\mathrm{K}}
\newcommand{\pc}{\mathrm{pc}}
\newcommand{\kB}{k_\mathrm{B}}
\newcommand{\au}{\mathrm{A.U}}
\newcommand{\Rin}{R_\mathrm{in}}
\newcommand{\Rout}{R_\mathrm{out}}
\newcommand{\Rstar}{R_\mathrm{\star}}
\newcommand{\Mstar}{M_\mathrm{\star}}
\newcommand{\Tstar}{T_\mathrm{\star}}
\newcommand{\rhoin}{\rho_\mathrm{in}}
\newcommand{\Tin}{T_\mathrm{in}}
\newcommand{\Knuext}{\kappa_\nu^\mathrm{ext}}
\newcommand{\Cnuext}{C_\nu^\mathrm{ext}}
\newcommand{\V}[1]{\boldsymbol{#1}}      % vector
\newcommand{\M}[1]{\mathbf{#1}}          % matrix

\newcommand{\etal}{\emph{et al.}\xspace}

\usepackage{amsmath,amsfonts,amssymb}
\usepackage{graphicx}
\usepackage{xspace}
\usepackage[colorlinks=true, allcolors=blue]{hyperref}
\usepackage[perpage]{footmisc}
\title{I2C@2024: An interferometry imaging contest in 2024}

\author[a]{Florentin Millour}
\author[a]{Mathis Houllé}
\author[a]{Jérémy Perdigon}
\author[b]{Julien Drevon}
\author[c]{Ryan Norris}
\author[c]{Rebecca Proni}
\author[a]{Alexis Matter}
\author[d]{Ferreol Soulez}
\author[d]{Eric Thiebaut}
\affil[a]{Laboratoire Lagrange, UniCA, CNRS, OCA, Nice, France}
\affil[b]{European Southern Observatory, Alonso de Córdova, 3107 Vitacura, Santiago, Chile}
\affil[c]{Department of Physics, New Mexico Institute of Mining and Technology, 801 Leroy Place, Socorro, NM 87801, USA}
\affil[d]{Univ Lyon 1, ENS de Lyon, CNRS, Centre de Recherche Astrophysique de Lyon,\\ UMR5574, F-69230, Saint-Genis-Laval, France}

\authorinfo{Further author information: (Send correspondence to F. Millour)\\F. Millour: E-mail: fmillour@oca.eu, Telephone: +33 489 15 03 59}

\begin{document} 
\maketitle

\begin{abstract}
Images in optical long-baseline interferometry have seen a boost in the recent years thanks to new techniques and recipes invented by the community. These images are more and more used for science interpretation and not only illustration, and their fidelity has improved significantly, thanks mainly to the increase in the number of telescopes used in interferometers. The focus today is to improve their reliability and dynamic range. With this contest, we follow up the quest introduced in 2004 of comparing the state of the art image reconstruction software for long-baseline interferometry. This is done in a festive way in the form of an imaging contest, where the organizers propose simulated datasets of targets, whose brightness distributions are meant to be blindly retrieved using various algorithms by the contestants. A prize is offered to the winner of the contest. This year is not different from previous ones and we proposed to the contestants tools to compare their reconstructed images with original images. These tools are now distributed, together with example datasets and images, enabling further tests at home of any image reconstruction tool.
\end{abstract}


%%%%%%%%%%%%%%%%%%%%%%%%%%%%%%%%%%%%%%%%%%%%%%%%%%%%%%%%%%%%
\section{INTRODUCTION}
\label{sec:intro}  % \label{} allows reference to this section

\begin{figure} [ht]
   \begin{center}
   \begin{tabular}{c} 
   \vspace{-3cm}\hspace{-1.5cm}
   \includegraphics[width=1.15\textwidth]{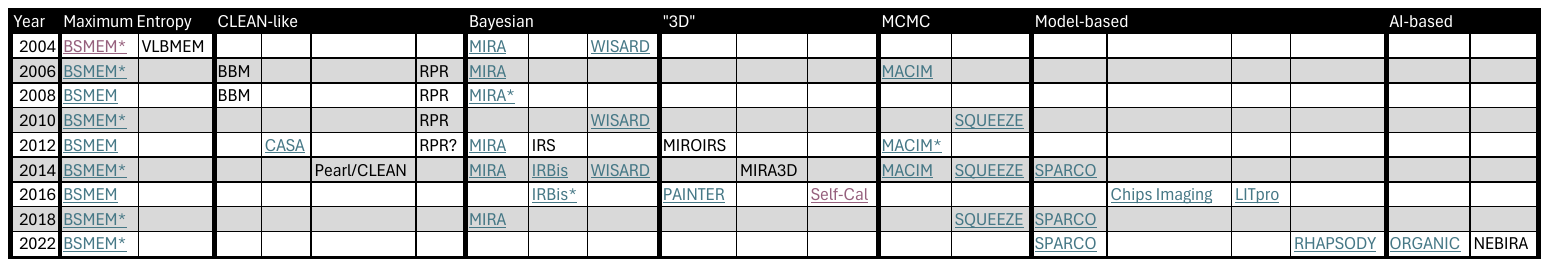}
   \vspace{-7cm}
	\end{tabular}
	\end{center}
   \caption[example] 
   { \label{fig:contests_digest} 
All the participating algorithms used in previous interferometry imaging contests. Winning algorithms are marked with an asterisk, and we tried as much as possible to group algorithms by family.}
   \vspace{1cm}
   \end{figure}

Long baseline interferometry contests have been organized since 2004 \cite{BC2004,BC2006,BC2008,BC2010,BC2012,BC2014,BC2016,BC2018,BC2022} and have enabled a comparison of algorithms' performances along their many editions. A total of 20 algorithms were used in the 18 years of previous contests. We provide in Fig.~\ref{fig:contests_digest} a digest of the algorithms running for the imaging contest, showing the prominent ones as well as the ones that may deserve a special attention. Among the winner algorithms (marked with an asterisk), we can find Maximum Entropy Methods, implemented in BSMEM\cite{2008SPIE.7013E..3XB}, Bayesian approaches to image reconstruction, like in MIRA\cite{Thiebaut2008}, or IRBIS\cite{2014AandA...565A..48H}, and statistical methods like the one implemented in MACIM\cite{BC2012}.
Alternate methods include attempts for 3D image reconstructions (including an implementation of self-cal in use with MIRA\cite{2011AandA...526A.107M}), mix of model-fitting and image reconstruction, with a special mention to SPARCO\cite{2014AandA...564A..80K} and RHAPSODY\cite{2022SPIE12183E..1OD}, as well as machine learning-based algorithms\cite{2020SPIE11446E..1UC}.

In the 2024 contest, we wished the contestants to address wavelength-dependent datasets on objects representative of typical imaging cases, namely a spiral nebula, that can be found in e.g. interacting binaries, and a multi-gapped disk, with planets embedded inside of different fluxes and spectral shape.

%%%%%%%%%%%%%%%%%%%%%%%%%%%%%%%%%%%%%%%%%%%%%%%%%%%%%%%%%%%%
\section{The contest data}

To produce the contest data, we tried as much as possible to use the JMMC tools to simulate objects and datasets.

\subsection{Spiral model}

\begin{figure} [ht]
   \begin{center}
   \begin{tabular}{c}
   \hspace{-0.5cm}
   \includegraphics[width=1.15\textwidth]{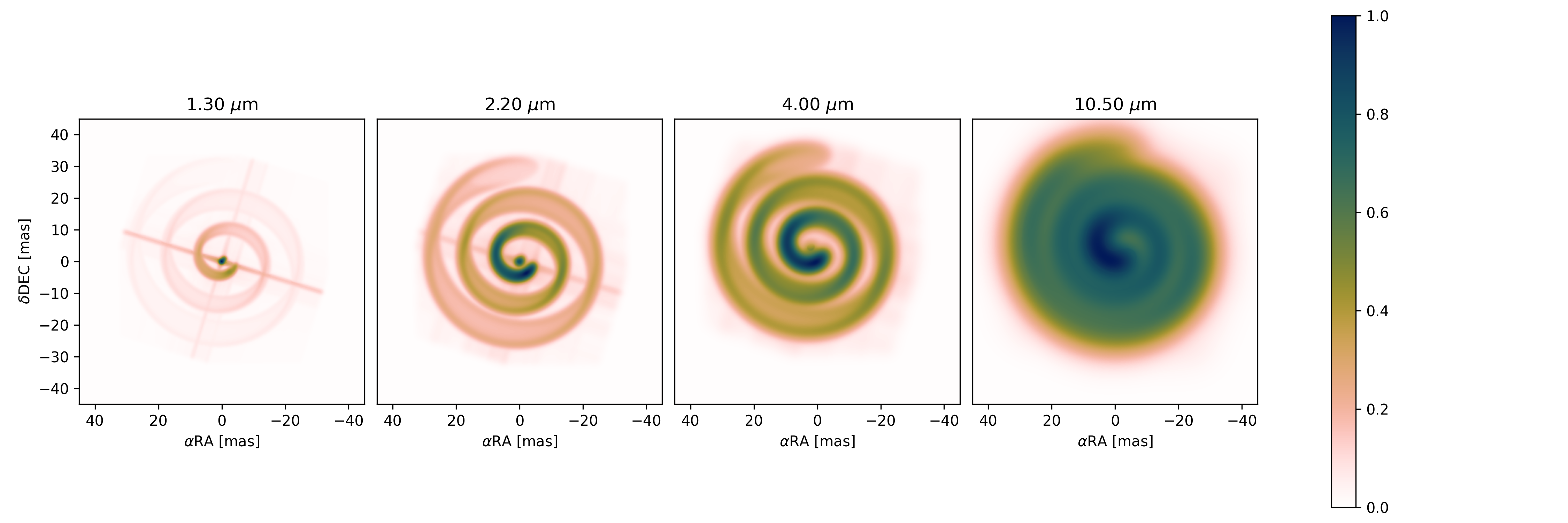}
	\end{tabular}
	\end{center}
   \caption[example] 
   { \label{fig:inputmodelspiral} 
Spiral model at selected wavelengths, showing the main structures we expect to see reconstructed by the contestants. Intensity scale is normalized to its maximum and displayed as the cubic root of the image intensity to make faint structures more visible.}
   \end{figure}

We used the AMHRA\footnote{\url{https://amhra.oca.eu/AMHRA/index.htm}} model of a spiral (derived from Millour et al. 2008\cite{Millour2008}) to produce an image cube in the H, K, L and N bands. The size was selected so that the first spiral turn is about 10\,mas in size, and has a total of 3 turns. All the parameters of the spiral are shown in Table~\ref{tab:spiral}, and the corresponding model images, convolved with the beam of the interferometer is shown in Figure~\ref{fig:inputmodelspiral}

\begin{table}
    \centering
    \begin{tabular}{c|c||c|ccccc}
        \hline
        \hline
        Type of spiral & rings        & Dust temperature & 1500\,K \\
        Filling factor & 20           & $q$ & 0.3\\
        Rounds & 3                    & Gap factor & 1.5 \\
        Opening angle & 40$^\circ$    & WR temperature & 70,000\,K\\
        Distance to first turn (size) & 10\,mas & OB temperature & 40,000\,K \\
        Inclination & 24$^\circ$      & Luminosity ratio of stars & 3 \\
        Rotation in spiral plane & 75$^\circ$ & Fraction of flux from the binary & 5 \\
        Distance to shocked material (hollow middle) & 4\,mas & Scale factor for pinwheel ($\propto$ flux) & $10^{-16}$ \\
        Binary separation & 1\,mas\\
        \hline
        \hline
    \end{tabular}
    \caption{Spiral model parameters as put into the AMHRA service.}
    \label{tab:spiral}
\end{table}

%\begin{table}
%    \centering
%    \begin{tabular}{c|c||c|ccccc}
%        \hline
%        \hline
%        Type of spiral & rings \\
%        Filling factor & 20 \\
%         Rounds & 3\\
%         Opening angle & 40$^\circ$\\
%         Distance to first turn (size) & 10\,mas \\
%         Inclination & 24$^\circ$\\
%         Rotation in spiral plane & 75$^\circ$\\
%         Distance to shocked material (hollow middle) & 4\,mas \\
%         Binary separation & 1\,mas\\
%         Dust temperature & 1500\,K\\
%         $q$ & 0.3 \\
%         Gap factor & 1.5 \\
%         WR temperature & 70,000\,K\\
%         OB temperature & 40,000\,K\\
%         Luminosity ratio of stars & 3\\
%         Fraction of flux from the binary & 5\\
%         Scale factor for pinwheel ($\propto$ flux) & $10^{-16}$\\
%         \hline
%         \hline
%     \end{tabular}
%     \caption{Spiral model parameters}
%     \label{tab:spiral}
% \end{table}

The AMHRA services allowed us to generate the spiral model as a cube of images for each set of wavelengths of interest. We fed these image cubes to ASPRO to get OIFITS files for each instrument, producing a realistic VLTI (u,v) coverage. We checked the "Compute OIFITS data", "Add error noise to the data", as well as the "Use instrumental and calibration error biases" to produce the OIFITS files of the contest. We verified using the Oimaging tool\footnote{\url{https://www.jmmc.fr/english/tools/data-analysis/oimaging/}}, especially using the MIRA \cite{Thiebaut2008} and SPARCO \cite{Kluska2014} implementations, that we could indeed reconstruct images that visually looked like the input models from the contest OIFITS files.

\subsection{Disk with planets model}

We simulated a disk of dust around a young A-type star $\left( \Tstar = 10 \ 000 ~ \K, \, \Rstar = 2.2 ~ R_\mathrm{\odot}, \, \Mstar = 2.8 ~ M_\mathrm{\odot} \right)$, using the ray tracing code from Perdigon et al. (in prep), to which we superimposed point sources representing planets. We aimed to produce a model that is similar to the circumstellar environment around a young Herbig Ae/Be star \cite{2016AandA...595A.112G}.

The ray tracing code computes synthetic images from the integration along rays of the circumstellar source function. In order to achieve the integration, the code requires the optical properties of the medium, namely the source function and the extinction coefficient, at each point in the circumstellar envelope. In the following, we describe the properties of the dust we used to compute the synthetic images of the disc. We then explain how we consequently injected the planets in the image.

\begin{figure}
    \centering
    \includegraphics[width=0.49\textwidth]{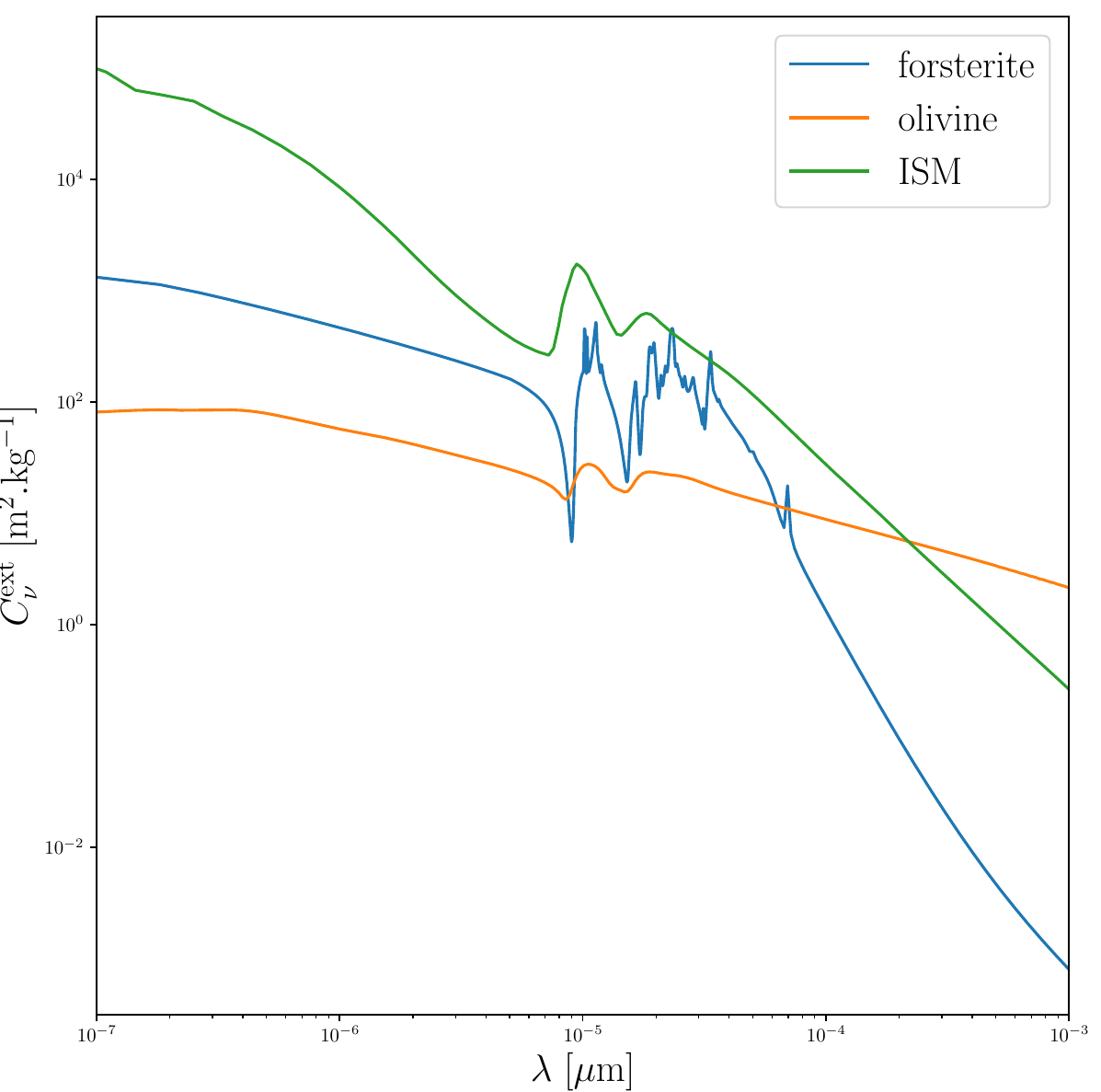}
    \includegraphics[width=0.49\textwidth]{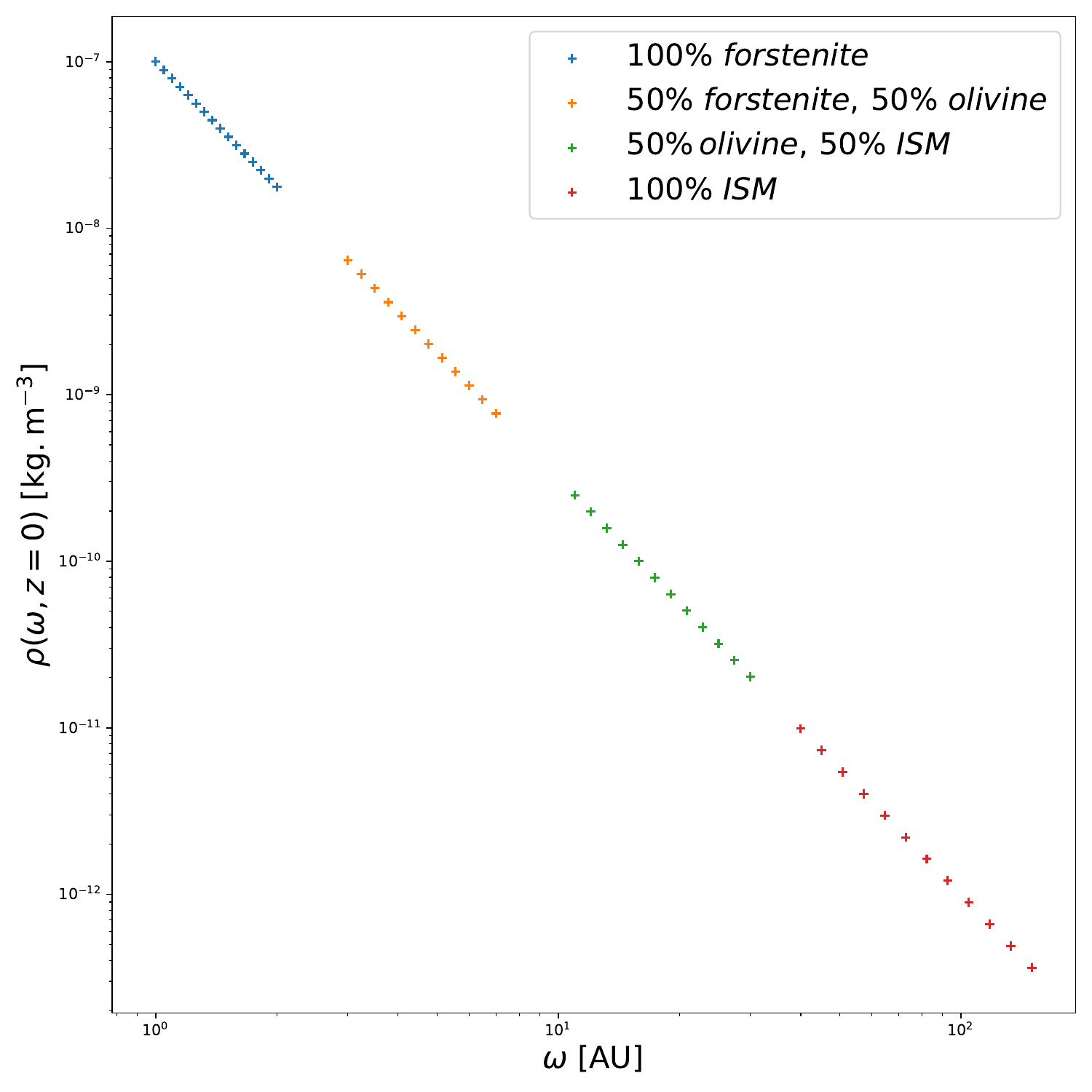}
    \caption{{\bf Left:} Extinction cross sections for the three different dust species used in the disk model. {\bf Right:} Density profile in the disk equatorial plane, colored by dust composition.}
    \label{fig:opac}
\end{figure}

The disk is made of four concentric rings separated by three gaps (Fig.~\ref{fig:opac}-Right). The smallest dust inner rim is set to $\Rin = 1 ~ \au$ while the largest outer radius is $\Rout = 150 ~ \au$. The disc is situated at a distance $d = 184 ~ \pc$. These values insure all components remain resolved by the VLTI instruments. The first gap ranges $2$ to $3 ~ \au$, the second ones spans from $7$ to $11 ~ \au$ and the last one extends from $30$ to $40 ~ \au$. The object is seen at an inclination of $i = 70 ~ \mathrm{deg}$ with respect to the pole.
 
The dust extinction coefficient $\Knuext$ is given by
\begin{equation}
    \Knuext = \Cnuext \, \rho
\end{equation}
where $\Cnuext$ are the dust extinction cross sections and $\rho$ is the dust density. The composition of the dust varies from one ring to another. We included this feature to reflect the decrease in dust crystallinity with increasing radius, due to associated changes in temperature, pressure and radiation. The composition of the four rings are (from the inner radius to outer radius): (i) $100 ~ \%$ forsterite, (ii) $50 ~ \%$ forsterite, $50 ~ \%$ olivine, (iii) $50 ~ \%$ olivine, $50 ~ \%$ ISM (silicate and graphite) and (iv) $100 ~ \%$ ISM. The extinction cross-sections, for each dust species, are shown in Fig.~\ref{fig:opac}-Left.

The density law is given by that of a disk hydrostatically supported in the vertical direction \cite{Carciofi2006}
\begin{equation}
    \rho = \rhoin \, \left(\frac{\omega}{\Rin}\right)^{-5/2} \, \exp{\left\lbrace-\frac{1}{2}\left(\frac{z}{H}\right)^2\right\rbrace}
\end{equation}
where $\omega$ is the equatorial radius, in the disk mid-plane and $z$ is the height with respect to the equatorial plane. $\rhoin$ is the dust density at the base of the disk and is arbitrarily set to $\rhoin = 10^{-7} ~ \mathrm{kg.m^{-3}}$. This value insures that the disk remains optically-thick to the observer. Finally, $H$ is the scale height of the disk and is set to be a power-law
\begin{equation}
    H = H_0 \, \left( \frac{\omega}{\Rstar}\right)^{13/10}
\end{equation}
with $H_0 = \left( \kB \, \Tstar \, \Rstar^3 / \left( m_{H_2} \, G \, \Mstar \right) \right)^{1/2}$. The exponent $13/10$ insures the disk does not flare too much so that the inner parts (such as the star and the inner radius) are not obscured by the outer rings.

For the source function $S_\nu$, we omit scattering and assume the dust emits like a black-body spectrum $S_\nu = B_\nu(T)$. The dust temperature is also chosen to be a power law, on the form
\begin{equation}
    T = \Tin \, \left( \frac{\omega}{\Rin} \right)^{-1/2}.
\end{equation}
We set the inner dust temperature $\Tin = 1500 ~ \mathrm{K}$ to be consistent with the usual dust evaporation temperature of silicate-based dust species. The temperature radial exponent $-1/2$ is that of a grey optically-tick and spatially flat disk \cite{Beckwith1999}.

For the substellar companions, we injected point sources with spectra extracted from the ExoREM model grid \cite{Charnay2018}. The atmospheric parameters and the relative astrometry of each companion are reported in Table~\ref{tab:companions}. Before injection, the spectra are convolved at the spectral resolutions of the instruments by a Gaussian filter\footnote{implemented by the \texttt{scipy.ndimage.gaussian\_filter} function}, and interpolated on the instruments' wavelength grids\footnote{with \texttt{numpy.interp}}.

\begin{figure} [ht]
    \begin{center}
        \begin{tabular}{c} 
            \includegraphics[width=\textwidth]{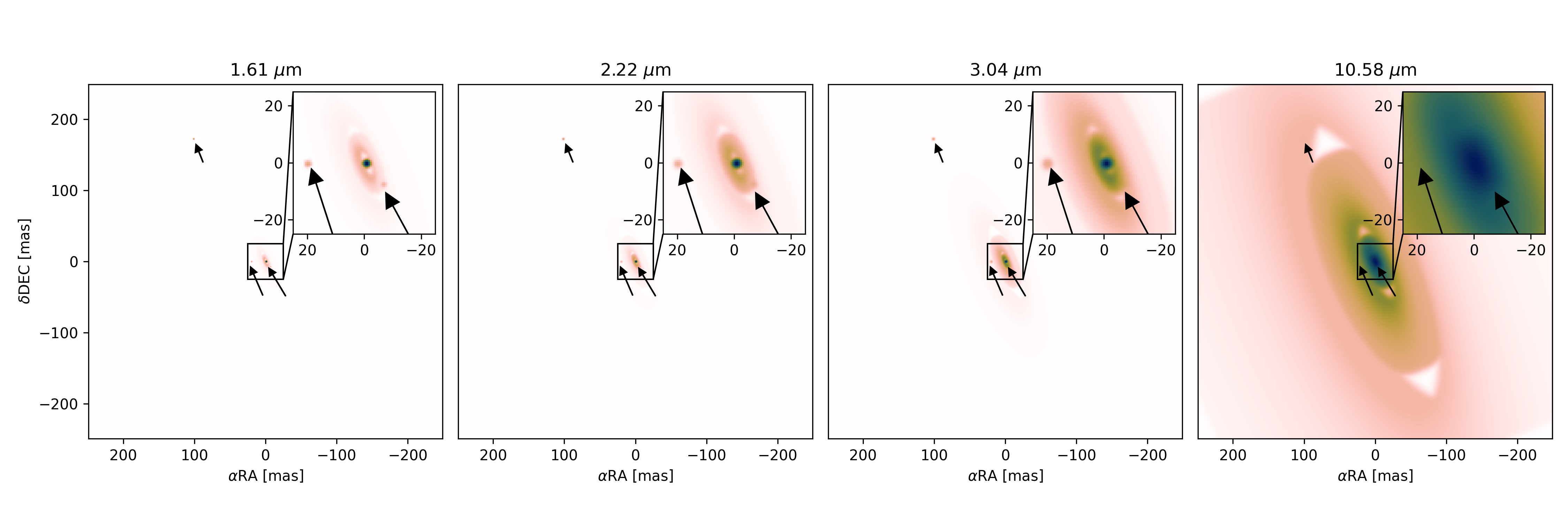}
	   \end{tabular}
    \end{center}
    \caption[example]{\label{fig:inputModelDisk} Disk model at selected wavelengths. The arrows mark the position of the three planets to be recovered. Notice the large scale of the image. The inset shows a zoom into the central region. Intensity scale is the cubic root of the image intensity to make faint structures more visible.}
\end{figure} 
\begin{figure} [ht]
    \begin{center}
        \begin{tabular}{c} 
            \includegraphics[width=\textwidth]{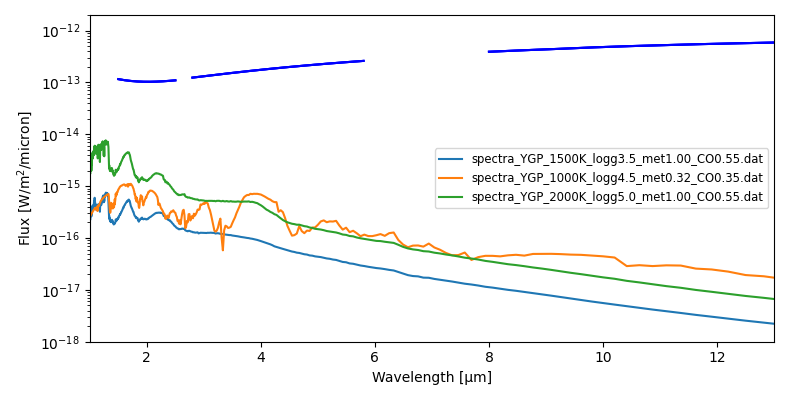}
	   \end{tabular}
    \end{center}
    \caption[example]{\label{fig:inputModelPlanets} Disk model spectrum (top) as well as injected planets spectra (bottom).
    }
\end{figure}

\begin{table}
\vspace{1cm}
    \caption{Injected substellar companions parameters.}
    \centering
    \begin{tabular}{c|ccccccc}
        Companion & Pos. X [mas] & Pos. Y [mas] & $T_{\mathrm{eff}}$ [K] & $\log g$ & [M/H] & C/O & $R$ [$R_\mathrm{Jup}$]\\
        \hline
        Planet 1 & -6.5 & -7.5 & 1500 & 3.5 & 1.0 & 0.55 & 2.0 \\
        Planet 2 & 20.5 & -0.5 & 1000 & 4.5 & 0.32 & 0.35 & 2.0 \\
        Brown dwarf & 101.5 & 171.5 & 2000 & 5.0 & 1.0 & 0.55 & 2.0 \\
    \end{tabular}
    \label{tab:companions}
\end{table}

In the same way as for the spiral model, we fed the model image cubes to ASPRO to get OIFITS files for each instrument, producing a realistic VLTI (u,v) coverage. We checked the "Compute OIFITS data", "Add error noise to the data", as well as the "Use instrumental and calibration error biases" to produce the OIFITS files of the contest. We verified using the Oimaging tool that we could indeed reconstruct images that visually looked like the input models from the contest OIFITS files.

The resulting dataset is one OIFITS file per instrument and infrared band. The datasets are chromatic, and have typical spectral resolutions, observables, noises and biases produced by each instrument. We posted the datasets on the OiDB database to make it available to anyone willing to participate to the contest this year\footnote{
\url{http://oidb.jmmc.fr/collection.html?id=6df579f8-752b-4536-b579-f8752bd536d0}}, set up a contest page providing general information about the contest\footnote{\url{http://fmillour.com/index.php/2024/03/25/2024-interferometry-imaging-contest/}}, and we provided and distributed largely a set of instructions\footnote{\url{https://docs.google.com/document/d/1F6mkPixTRgC6VTxyzoGaWKPQaFroEqFRY79NnWsLaFc/edit?usp=sharing}} that we detail in the next section

%%%%%%%%%%%%%%%%%%%%%%%%%%%%%%%%%%%%%%%%%%%%%%%%%%%%%%%%%%%%
\section{Instructions to the contestants}

On the general webpage presenting the contest, we provided some context to the targets procided in the datasets: "we have two targets observed with all the VLTI instruments (PIONIER, GRAVITY, and the two L- and N-band arms of MATISSE).
– The first one is a hot star with an environment,
– The second one is a young star where there is a suspicion of a companion.
One of them is an easy target, so we will expect to see details of the structures in the images. The other one on the contrary is a rather difficult target, so we expect to see the prominent features and structures in it."

We publicised the calendar of the contest in the distributed set of instructions in the following: 1$^{\rm st}$ of March 2024: contest announcement, take contact with interested teams ; mid-March, datasets distribution ; May 31: the teams send their results to the organizers, comparison of results with input images, start of proceeding writing. At about the same time or earlier: distribution of the codes to establish the score ; presentation of results at SPIE in June.

The contestants obtained oifits files for several objects that were produced using simulations of observations on the Very Large Telescope Interferometer, with all available instruments (i.e. one oifits file per instrument / band). The datasets were chromatic, and had typical spectral resolutions, observables, noises and biases produced by each instrument.

We expected from the contestants image cubes that contained 3D images: 2D images plus one additional wavelength dimension.
Images scales were supposed to be provided using the CRPIX1, CRVAL1, CDELT1, CUNIT1, CRPIX2, CRVAL2, CDELT2, CUNIT2 (spatial scales) and CRPIX3, CRVAL3, CDELT3, CUNIT3 (wavelength scale)\footnote{see here for instructions how to set it up: 
\url{https://github.com/JMMC-OpenDev/aspro-doc/blob/main/aspro-doc.md##user-defined-model}}.

We also provided general rules as: if no image was provided by the contestant for a given wavelength/band, it would have been replaced by a uniform image. We warned the contestants that such a procedure would have heavily penalized the score, this was to strongly encourage contestants to provide images at all wavelengths.

We finally provided a basic description of the metrics that would be used for the image comparison, and provided some FAQ treating the questions from the contestants that arrived along the way.

%%%%%%%%%%%%%%%%%%%%%%%%%%%%%%%%%%%%%%%%%%%%%%%%%%%%%%%%%%%%
\section{Contest entries}

We contacted the whole interferometry community for the contest through the OLBIN\footnote{\url{olbin@univ-grenoble-alpes.fr}} mailing list and obtained 15 answers from various teams, that expressed interest in the contest. Out of these, we received two entries that both contained chromatic images of the objects of interest. We present them here and after we compare their entries with the input images.

The few contributions this year may be a symptom of a fading interest of the experts in the development of new imaging algorithms, or the fact that this year's contest was focused on producing realistic instrument datasets, including calibration biases, that are indeed usually much harder to tackle than simpler simulated data that can be found in many former imaging contests. Another possibility is that simply most of the available image reconstruction tools (including the most popular ones) are still not ready for real interferometry datasets (heavily multispectral, with tons of systematics), a situation that calls for new developments and adaptations of algorithms.

Nevertheless, the two entries received were of very high quality and both neatly managed to reproduce the main structures of the input objects, meaning that future contests may focus more on dynamic range exploration, as we tried to experiment for the second object.

As a side note, none of the participants followed closely the provided guidelines for images submission, especially for the wavelength scale, meaning that a more thorough explanation together with an example script may be provided for future editions of the contest. For this contest, we adapted the wavelengths scales for one participant, and asked one of the particpants to submit a new contribution with a sorted out wavelength scale.

\subsection{J. Drevon entry using Mira}

\begin{figure} [ht]
   \begin{center}
   \begin{tabular}{c} 
   \includegraphics[width=\textwidth]{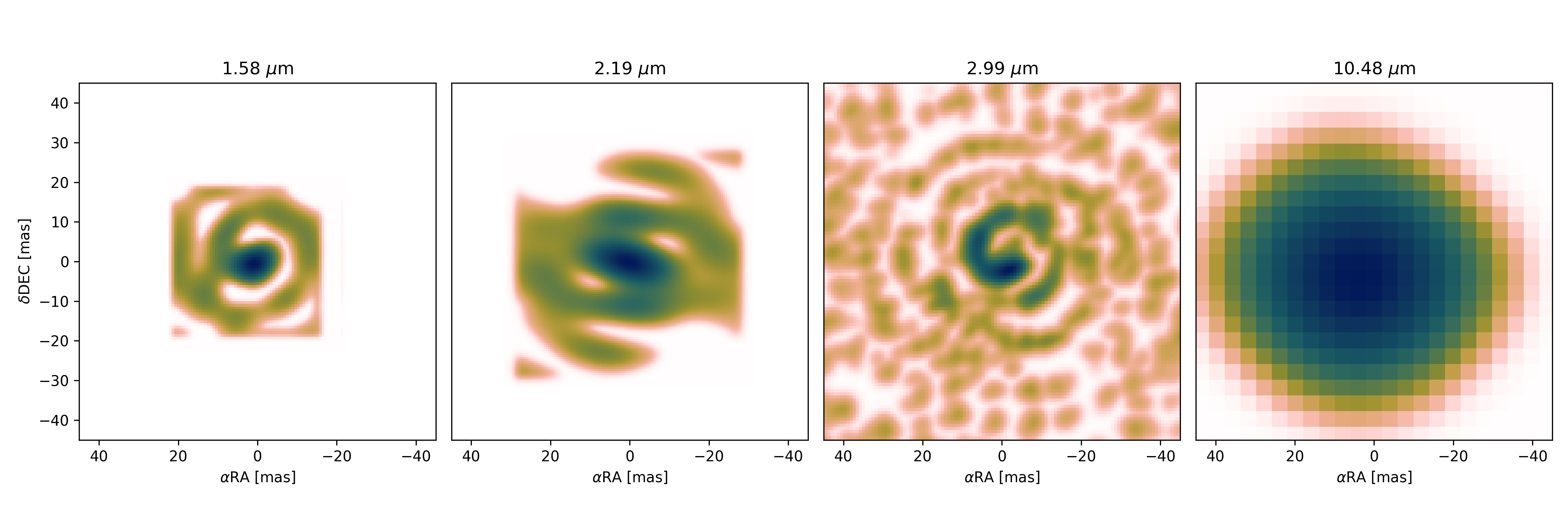}\\
   \includegraphics[width=\textwidth]{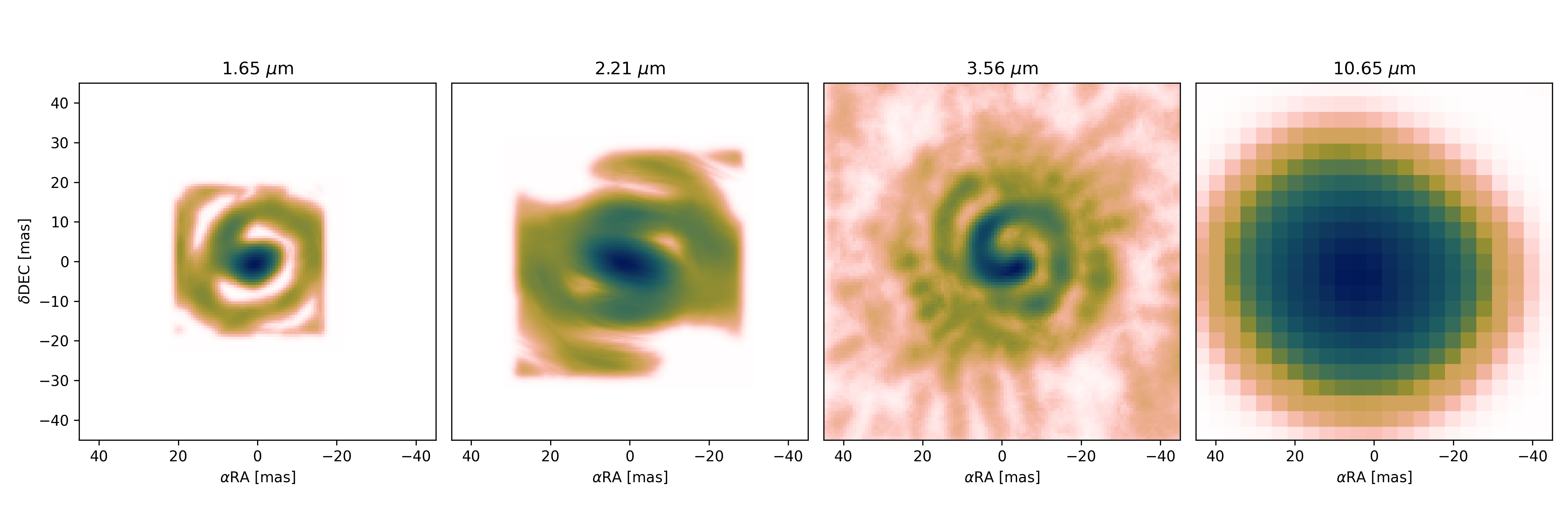}
	\end{tabular}
	\end{center}
   \caption[example] 
   { \label{fig:drevonSpiral} 
Spiral entry from J. Drevon. {\bf Top:} images at one selected wavelength. {\bf Bottom:} the median image of all entry wavelengths. Image scale is the cubic root of the intensity to make faint structures more visible.}
   \end{figure}

J. Drevon sent a set of input images cubes for both objects, one input file for each band. Figures~\ref{fig:drevonSpiral} and \ref{fig:drevonDisk} show on the top one single wavelength image for each band, while the bottom set of images shows the median of all wavelengths, that has a neat tendency to wipe out single-wavelength non-stationary artifacts in the images.

These entry images were interpreted by J. Drevon in the following way:

\subsubsection{Spiral}

\paragraph{PIONIER H-band:}
The image reconstructed with MiRA using the data from the PIONIER instrument, reveal a beginning of spiral formation close from the central source. This spiral is induced by a companion that reshape the circumstellar environment. The shape of the spiral arms seems to be wavelength dependent by comparison with the data from the following instruments.

\paragraph{GRAVITY K-band:}

The image reconstructions done using the data from GRAVITY reveal and confirm the previous image reconstruction, we observe an elongated structure with a beginning of a deformation in the circumstellar environment. We should take care about possible artifacts in the image reconstruction especially far from the central source.

\paragraph{MATISSE L-band:}

This time the image reconstruction seems to reconstruct centro-symmetric spiral arms. It is hard to say if the central source is resolved or if it is the product of artifacts in the image reconstruction process. Looking at the different images reconstructed for each spectral channel, we mainly resolve what might be a central source with a spiral arm. The spiral arm seem to have once again strong non-homogeneity in the spiral arms.

\paragraph{MATISSE N-band:}
For the N-band data of the VLTI/MATISSE instrument, it hard to resolve more than a close spherical-like Gaussian shape.

\subsubsection{Disk}
   
   \begin{figure} [ht]
   \begin{center}
   \begin{tabular}{c} 
   \includegraphics[width=\textwidth]{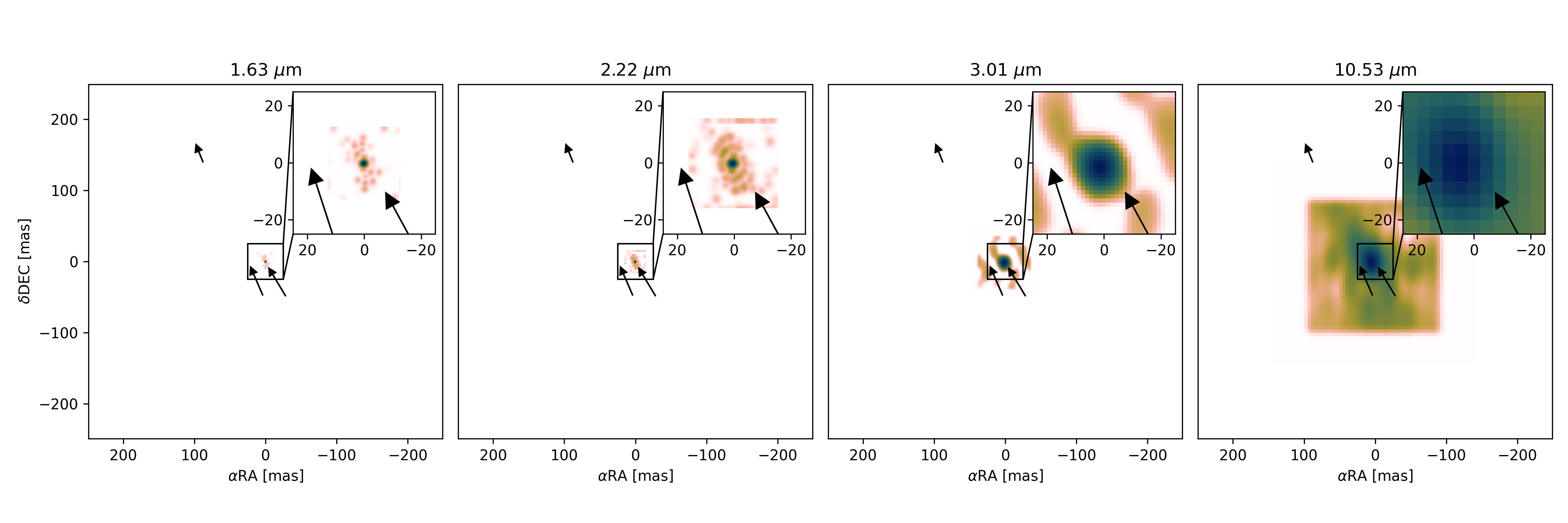}\\
   \includegraphics[width=\textwidth]{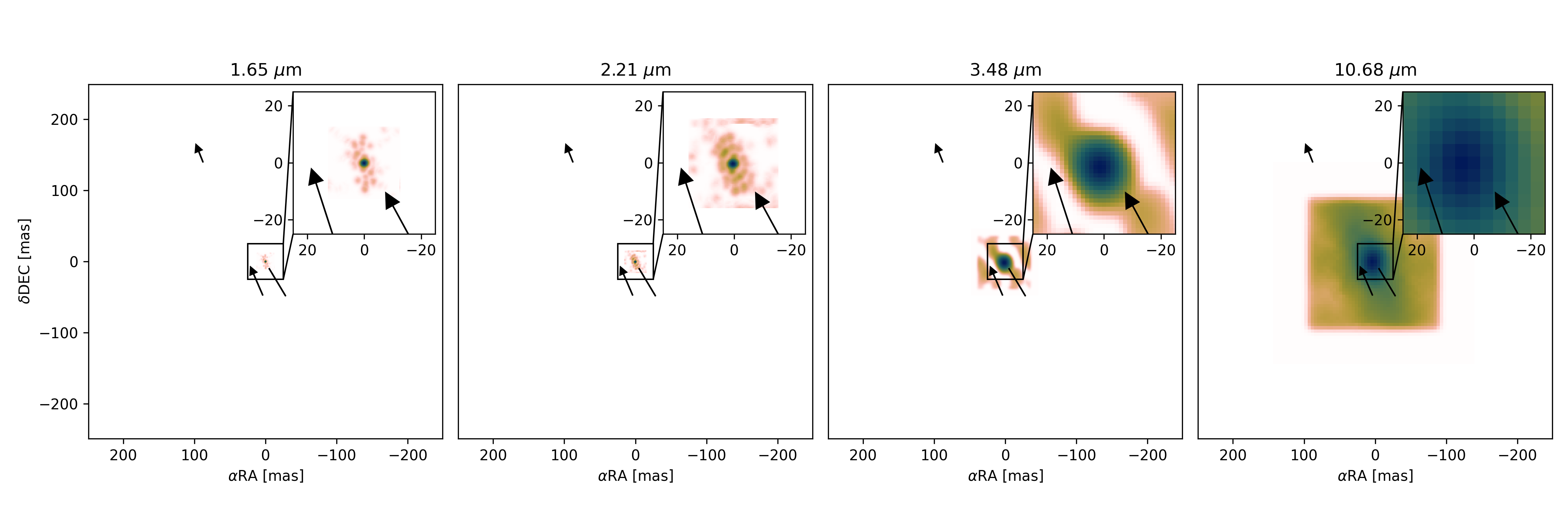}
	\end{tabular}
	\end{center}
   \caption[example] 
   { \label{fig:drevonDisk} 
Disk entry from J. Drevon. {\bf Top:} a selected wavelength in the image cube. {\bf Bottom:} the median of all entry wavelengths. Image scale is the cubic root of the intensity to make structures more visible.
}
   \end{figure} 

\paragraph{PIONIER H-band:}

Image reconstruction procedure using the VLTI/PIONIER data in H-band only show an elongated structure filed by artifacts with a resolved central source.

\paragraph{GRAVITY:}

The image reconstruction procedure using the VLTI/GRAVITY data does not allow us to describe more the geometry of the object than what  have been done using PIONIER. We only detect an elongated structure filled by artifacts.

\paragraph{MATISSE L-band:}

With MATISSE we clearly observe a central source with very small elongation. It is not clear if the faint arms present around the central source are real or not. To be conservative on the conclusion, I decided to rank those structures as mostly artifacts.

\paragraph{MATISSE N-band:}

In N-band it was hard to reconstruct good images, however we can observe an elongated central source with some deviation from centro-symmetry. The central source seems to be embedded inside a thermal background. As previously mentionned in LM-band, it is hard to tell if the structures and the overdensity close to the central source in the thermal background are rather artifacts or real structures.

\subsection{R. Norris using SQUEEZE + OITOOLS.jl}

R. Norris used a combination of SQUEEZE and OITOOLS.jl to reconstruct wavelength-dependent images of the contest objects. He sent as well image cubes, one for each wavelength. He provided the following description prior to the evaluation process:

\subsubsection{Spiral}

\begin{figure} [ht]
   \begin{center}
   \begin{tabular}{c} 
   \includegraphics[width=\textwidth]{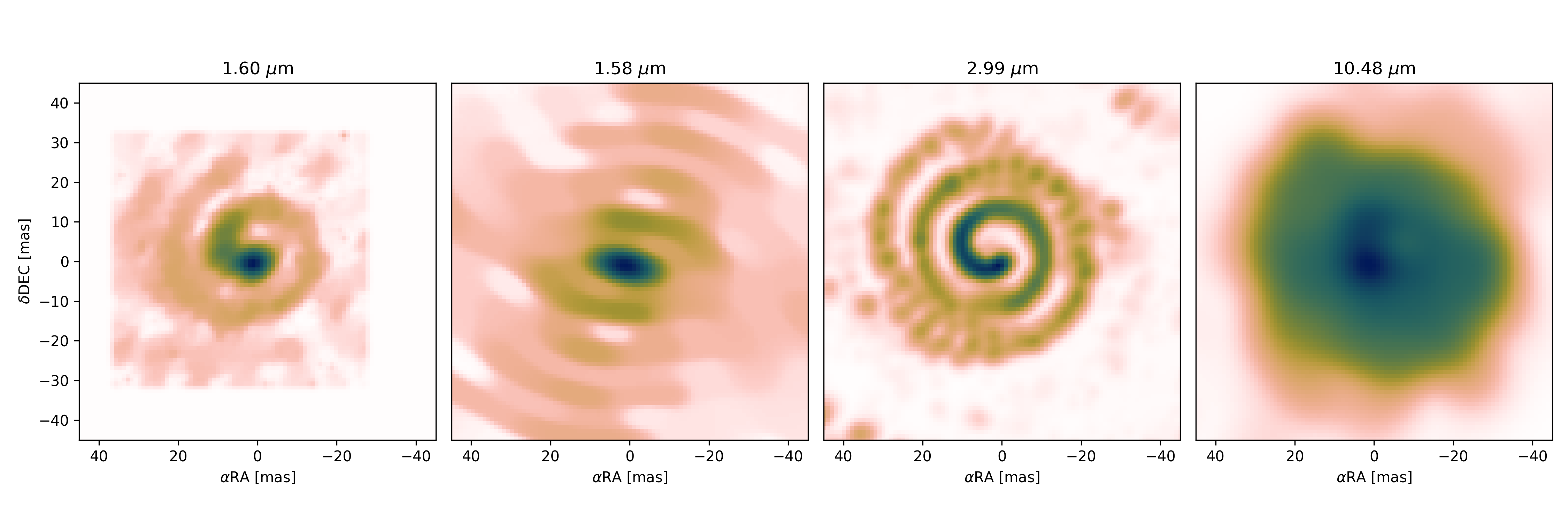}\\
   \includegraphics[width=\textwidth]{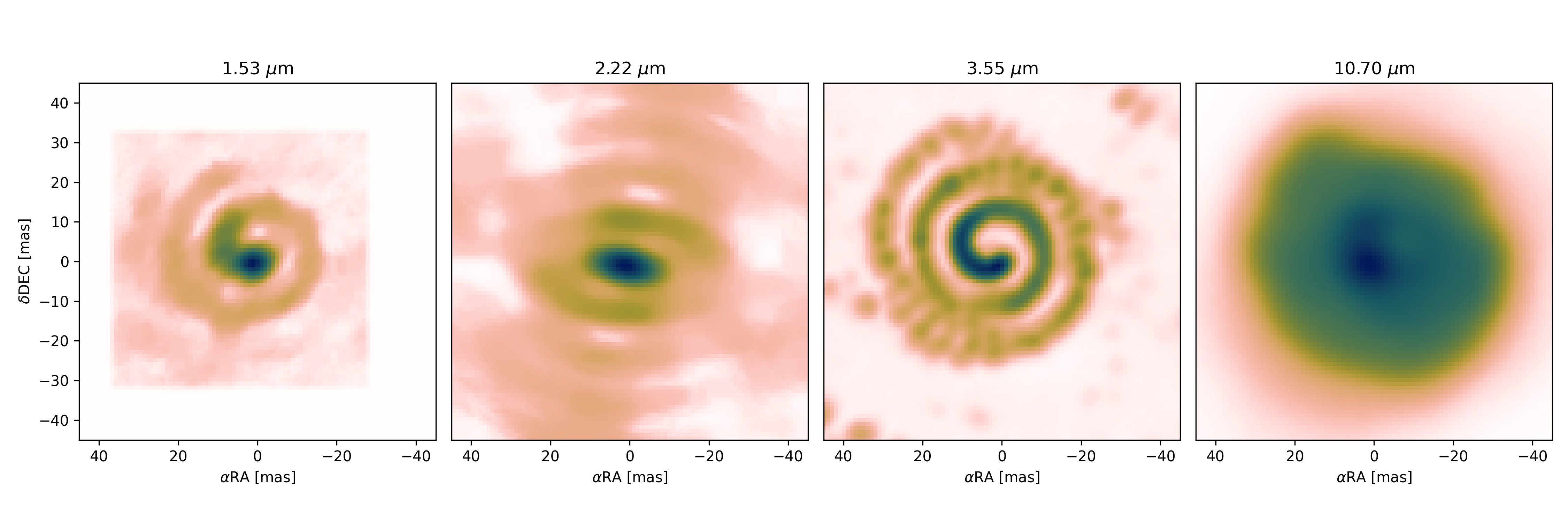}
	\end{tabular}
	\end{center}
   \caption[example] 
   { \label{fig:norrisSpiral} 
Spiral entry from R. Norris. Top is a selected wavelength, bottom is the median of all entry wavelengths. Image scale is the cubic root of the intensity to make structures more visible.}
\end{figure} 
   
Object 1 has the appearance of a Wolf Rayet system similar to WR104. The winds between the two stars in the system collide, compressing gas in a shock that cools to become dust. As seen in the PIONIER and MATISSE data, this forms a pinwheel  due to binary motion. In the GRAVITY image we see evidence of successive dust shells from past wind collision events. This also appears within the longer wavelength MATISSE images, which have a prominent feature from a denser region, similar to what was recently observed by JWST in WR104.

\subsubsection{Disk}
   
   \begin{figure} [ht]
   \begin{center}
   \begin{tabular}{c} 
   \includegraphics[width=\textwidth]{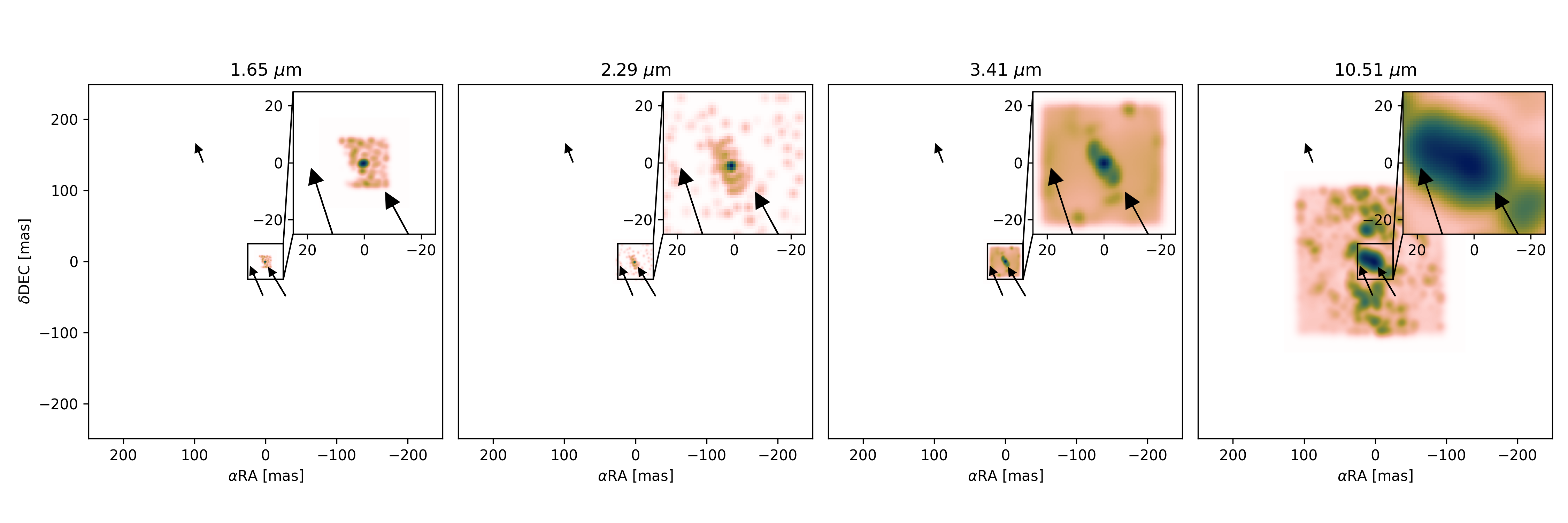}\\
   \includegraphics[width=\textwidth]{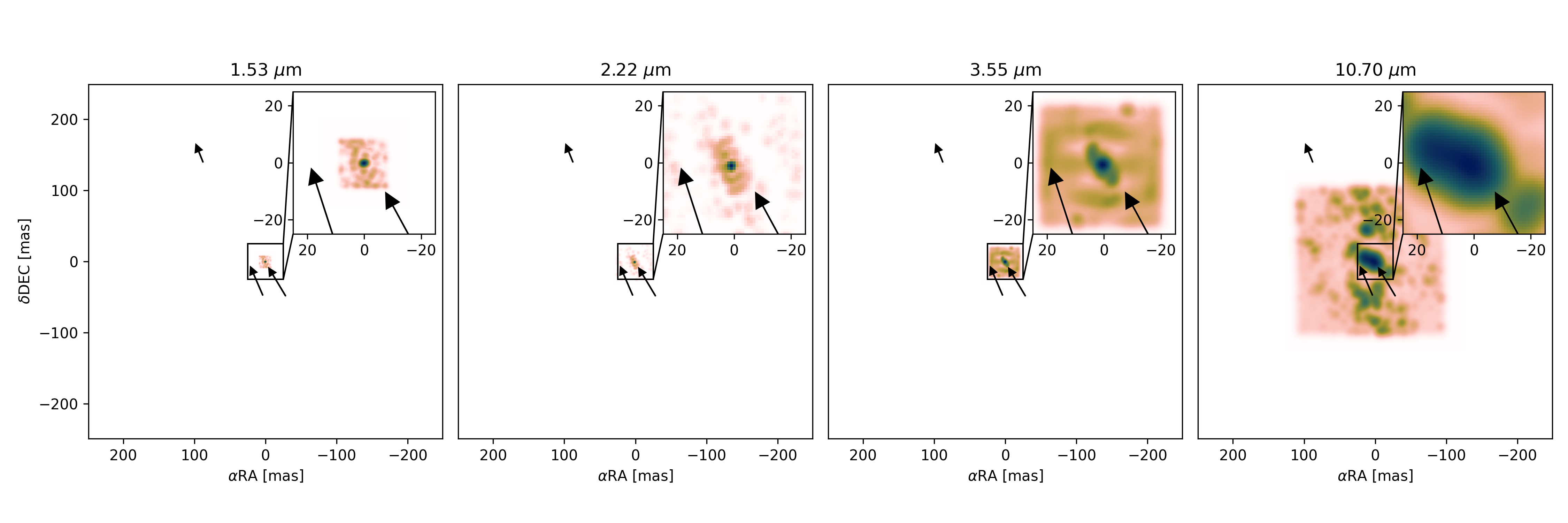}
	\end{tabular}
	\end{center}
   \caption[example] 
   { \label{fig:norrisDisk} 
Disk entry from R. Norris. Top is a single selected wavelength, bottom is the median of all entry wavelengths. Image scale is the cubic root of the intensity to make structures more visible.
}
   \end{figure} 
   
Object 2 is a region around a young star with a possible planet. In the PIONIER data, we see the central region as well as some denser areas in the outer disk. The image from  the GRAVITY data shows some more complex structures within the disk, though this is visible using a square root intensity scale in the image. The lower wavelength MATISSE image has a spiral appearance, which has been interpreted as evidence of a planet in some systems. The longer wavelength MATISSE image has the central bright central region once more and a feature to the NW. 

\section{How did we compute the contest scores}

Image comparison is a sensitive topic, especially for interferometric image reconstruction! 
To provide a fair comparison of images, it is important that the computed score does not depend on irrelevant changes. For instance, the image metric should be insensitive to some geometrical transform such as translation or to scaling: a multiplication of the brightness by some positive factor which does not affect the shape of the object. 
In addition, as we did not gave to the contestant any guidance about the size of the pixels in the reconstruction, the effective resolution achievable by the instrument must be taken into account when comparing a true image $\V{z}$ (with potentially an infinitely high resolution) to a restored image $\V{x}$. Otherwise, on pixel-wise comparisons, the slightest displacement of sharp features would lead to large loss of quality (according to the metric) whereas the images may look very similar at a lower and more realistic resolution.

This problem of assessing the quality of restored images has already been extensively studied in Gomez \etal \cite{Gomes2017} that based the evaluation of the best merit function on experts' poll and comparison of methods. Following its prescription, we use the L1 norm to compare the images as it seems to be the one that best match both the global minimum relative to the input image, and the human assessment. This L1 norm is  normalized by the reference image total flux. The resulting score between two images $\V{x}$ and $\V{y}$ can be described as follows:
\begin{equation}
    L(\V{x}, \V{y}) = \frac{ \sum_{k\in \textrm{pixels}}\left| x_k - y_k\right|}{\sum_{k\in \textrm{pixels}}\left| y_k\right|}
\end{equation}\label{eq:l1norm}
where the reference image $\V{y}$ is the true image $\V{z}$ blurred to a prescribed spatial resolution. To keep this metric insensitive to  translation and scaling, it is minimized by shifting the input image by $\V\delta=(\delta_{\mathrm{x}}, \delta_{\mathrm{y}})$ and scaling its flux by $\alpha$ for each wavelength. The overall multispectral score then writes:
\begin{equation}
S(\V{x}) = 1 - \frac{1}{N_\lambda}\sum_\lambda \min_{\alpha, \V\delta} L\left(\alpha\,\M{T}_\delta\cdot \V{x}_\lambda, \V{y}_\lambda \right)\,,
\end{equation}\label{eq:score}
where $\M{T}_{\V\delta}$ is an operator translating the image by $\V\delta=(\delta_{\mathrm{x}}, \delta_{\mathrm{y}})$ and $N_\lambda$ is the number of spectral channels in $\V{x}_\lambda$ and $\V{y}_\lambda$, the reconstructed and reference images in spectral channel $\lambda$. Since there are no closed-form expression of the best scaling factors $\alpha$, and best translation $\V\delta$ the score is numerically optimized using a bisection method.

In addition, the contestant submission does not necessarily match the model image in terms of field of view, pixel scale and wavelength scale.  
To alleviate these issues, we do the following assumptions and preprossessings:
\begin{itemize}
    \item during translation, we assumed the pixels outside of the field of view to be equal to zero,
    \item we generate the model image with a high spatial resolution and spatially resample the contestant image using bilinear interpolation to the resolution of the model image,
    \item similarly we resample the image spectrally to match the wavelength scale of both the model and the contestant's ones.
\end{itemize}

We produced scripts to implement this score computation, both in Julia and python, with examples snippets\footnote{ \url{https://github.com/JMMC-OpenDev/ImageMetrics} }. This allows anyone to run them on their side to test their own datasets as well as the 2016 interferometry imaging contest datasets.

\section{Results}

\begin{figure} [ht]
   \begin{center}
   \begin{tabular}{c} 
   \includegraphics[width=0.48\textwidth]{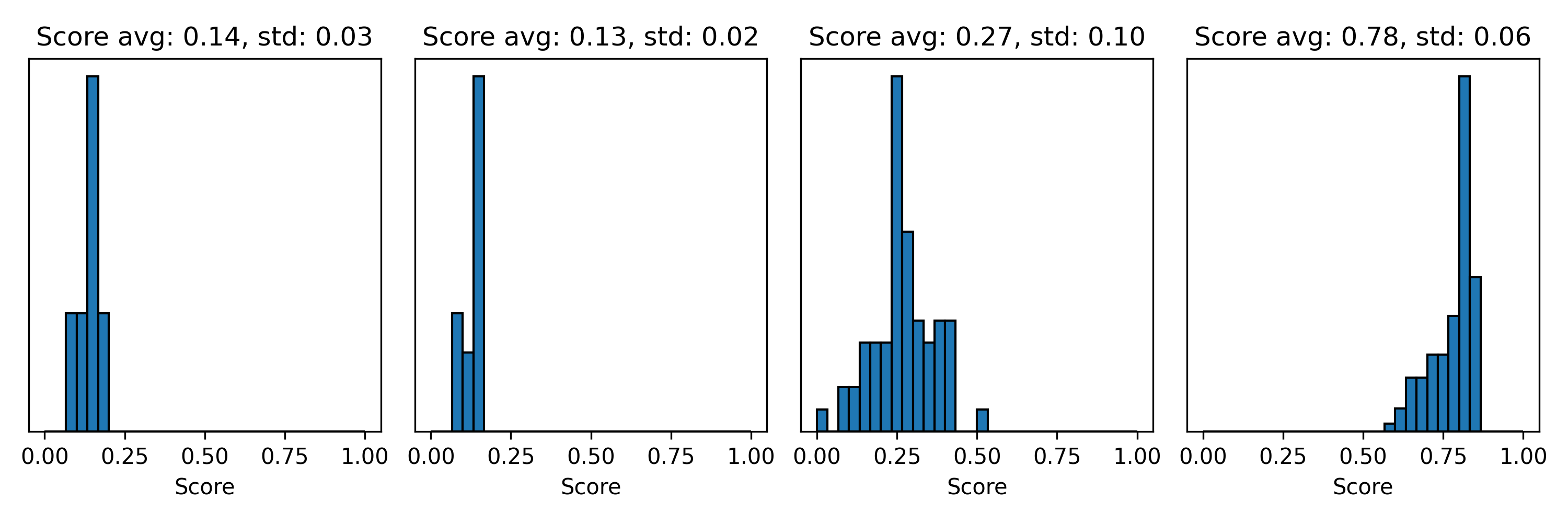}
   \includegraphics[width=0.48\textwidth]{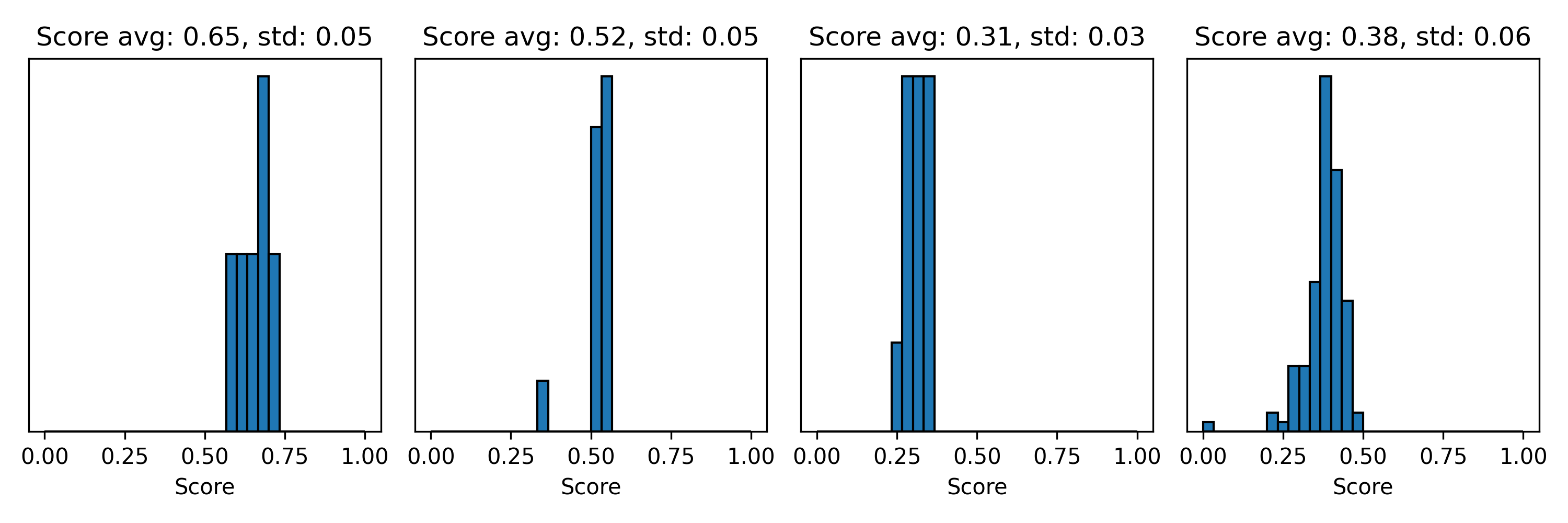}
	\end{tabular}
	\end{center}
   \caption[example] 
   { \label{fig:drevonScore} 
Score calculated with the recipe described above for the J. Drevon entries. The first 4 panels are for the spiral object, while the 4 later panels are for the disk object.
}
\end{figure} 

\begin{figure} [ht]
   \begin{center}
   \begin{tabular}{c} 
   \includegraphics[width=0.48\textwidth]{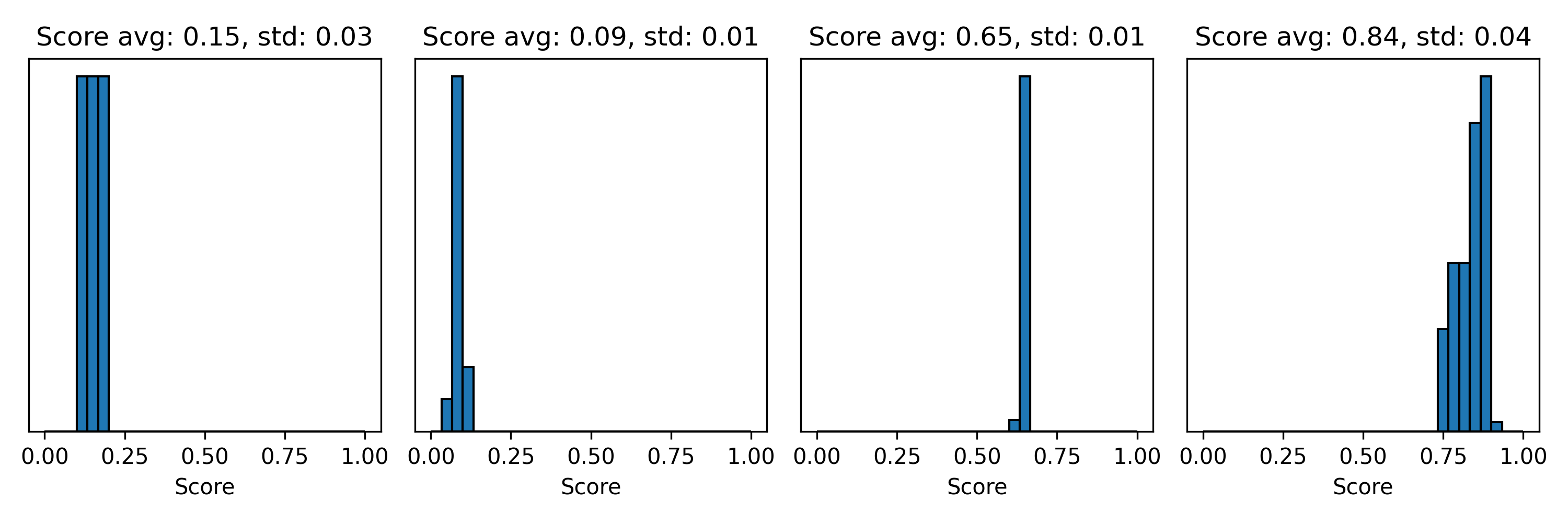}
   \includegraphics[width=0.48\textwidth]{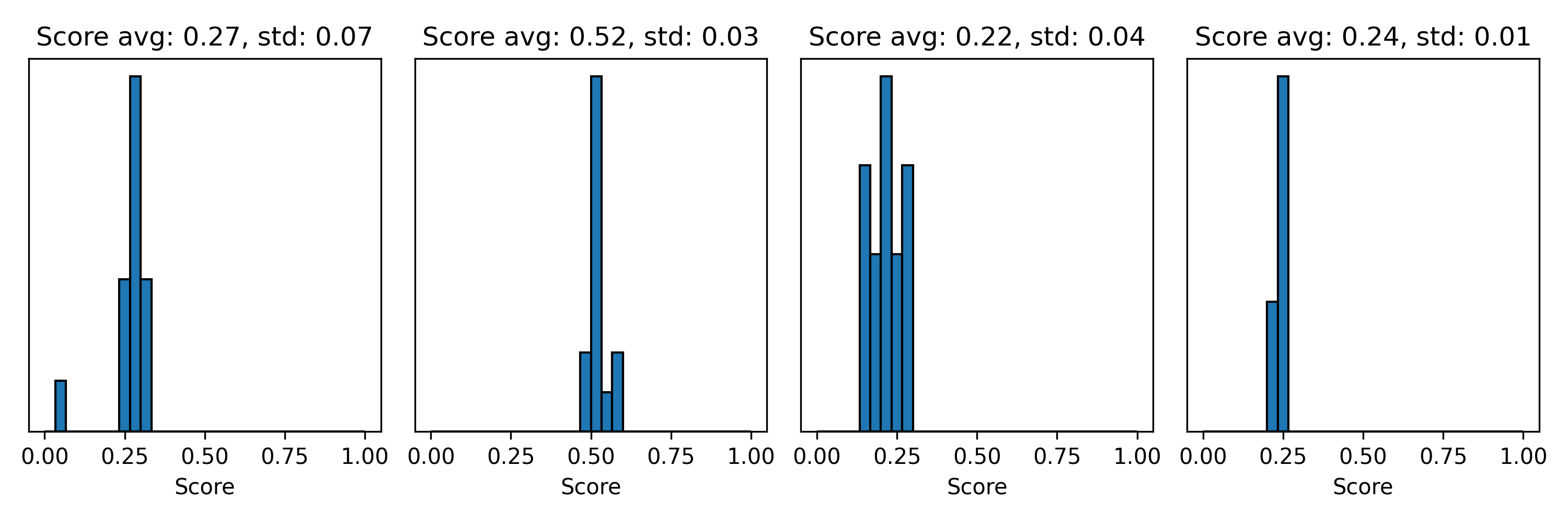}
	\end{tabular}
	\end{center}
   \caption[example] 
   { \label{fig:norrisScore} 
Score calculated with the recipe described above for the R. Norris entries. The first 4 panels are for the spiral object, while the 4 later panels are for the disk object.
}
\end{figure} 
   
All images of the input model and the entry images shown in figures \ref{fig:inputmodelspiral}, \ref{fig:inputModelDisk}, \ref{fig:drevonSpiral}, \ref{fig:drevonDisk}, \ref{fig:norrisSpiral}, and \ref{fig:norrisDisk}, are convolved to $\frac{\lambda}{4\,B_\text{max}}$ and are plot at the same scale, easing a visual comparison.

Wavelength-histograms of scores were computed for all participants and each image (see Figures~\ref{fig:drevonScore} and \ref{fig:norrisScore}). They all show peaked distributions, making us confident that the mean scores are indeed comparing relevant features in the (different for each wavelength) images, and not only tracing wavelengths-by-wavelength shifts or wavelength-dependent flux variations.

We compile the mean scores for each band here in Table~\ref{tab:final_scores}. The total score is just the sum of all scores, except the K-band spiral one, where both participants provided images that clearly do not look like the input image (keeping the K-band spiral would not change the final result).

\begin{table}
    \centering
    \begin{tabular}{c|cccc|cccc}
        Participant & Spiral H & Spiral K & Spiral L & Spiral N & Disk H & Disk K & Disk L & Disk N \\
        \hline
        J. Drevon & 0.14 & 0.13 & 0.27 & 0.78 & 0.65 & 0.52 & 0.31 & 0.38 \\
        R. Norris & 0.15 & 0.09 & 0.65 & 0.84 & 0.27 & 0.52 & 0.22 & 0.24 \\
        \hline
    \end{tabular}\\
    .\\
    .\\
    \begin{tabular}{c|cccc|cccc}
        Participant & Spiral score & Disk score & Total score \\
        \hline
        J. Drevon & 1.19 & 1.86 & 3.05 \\
        R. Norris & 1.64 & 1.25 & 2.89 \\
        \hline
    \end{tabular}
    \caption{Computed scores as in eq.~\ref{eq:score}}
    \label{tab:final_scores}
\end{table}

As a consequence, the contest organizers award the 2024 imaging contest prize to Julien Drevon for his nice images of both the disk and the spiral. We also wanted to make a special mention to Ryan Norris, that provided neat images of both objects and had a score close-enough to the best score to have it mentioned.

\section{Conclusion}

This year's contest showed nice entries with chromatic contributions that both showed very interesting images of the model images. We introduced example scripts and used a method to compare the images irrelevant of the input data, a method that provides a better way to evaluate whether the images match the input ones, thus not relying on the underlying datasets noises or particularities.

The small participation this year may be a sign that imaging softwares are not yet ready to handle real datasets, i.e. ones that contain large systematic errors (not necessarily accounted for in the oifits files) and massively multispectral information.

We may for the next editions of the imaging contest to change the agenda so that participants may have more time to produce their entries. Another focus may be on high-dynamic range imaging as was attempted this year for the disk model.

\acknowledgments % equivalent to \section*{ACKNOWLEDGMENTS}       
 
We warmly thank the SPIE interferometry session organizers, S. Sallum and J. Sanchez-Bermudez, for their support in the organization of the imaging contest this year.

This contest has been organized using funding of several projects and the community contribution of several tools: we warmly thank the JMMC for providing observation preparation and post-processing tools that were used for the preparation of this contest. We are also grateful for two French’s Agence Nationale de la Recherche projects: the project MASSIF (ANR-21-CE31-0018), and the project ExoVLTI (ANR-21-CE31-0017) that contributed to the preparation of the contest.

% References
\bibliography{report} % bibliography data in report.bib
\bibliographystyle{spiebib} % makes bibtex use spiebib.bst

\end{document}